# Extend the lifetime of wireless sensor networks by modifying cluster-based data collection

**Nahid Ebrahimi \*,  Ali Taghavirashidizadeh, Seyyed Saeed Hosseini**

1. Department of Technical Support, Isfahan Science & Technology Town, Isfahan, Iran, ebrahimi-n@istt.ir
2. Department of Electrical and Electronics Engineering, Islamic Azad University, Central Tehran Branch(IAUCTB), Tehran, Iran, ali.taghavi.eng@iauctb.ac.ir
3. Department of Electrical and Electronics Engineering, Islamic Azad University, Central Tehran Branch(IAUCTB), Tehran, Iran, saeedhosseini403851@gmail.com

**Abstract**

Wireless sensor networks have significant potential to increase our ability to view and control the physical environment, but the issue of power consumption in these networks has become an important parameter in their reliability and since in many applications of networks. Wireless sensor needs to guarantee end-to-end quality parameters, support for quality of service in these networks is of utmost importance. In wireless sensor networks, one of the most important issues is the lifetime of each network, which is directly related to the energy consumption of the network sensors. Increasing network lifetime is one of the most challenging requirements in these types of networks. This paper presents a clustering approach to reduce power consumption in wireless sensor networks, which is the most effective method for scalability and energy consumption reduction in sensor networks. The simulation results show that the proposed algorithm can reduce the energy consumption of the wireless sensor network and dramatically increase the lifetime of the network.

## 1. Introduction

Wireless sensor networks are one of the most important tools for acquiring information and understanding the environment that has led to extensive research. Wireless sensor networks consist of hundreds or thousands of nodes that are randomly distributed in remote or dangerous areas. The main task of these nodes is to collect information from the environment in which they are located. In fact, these nodes are capable of gathering information from these areas that are otherwise impossible to obtain or require a great deal of time and expense. Each sensor consists of a sensor, a computing unit, a memory and a wireless communication unit with a limited range of communication, all of which require some sort of energy source to continue their work. Despite advances in these types of networks, users and applicants of these networks are demanding an acceptable level of reliability in these systems, with energy consumption in the sensor network being the most important challenge in increasing reliability. There are a large number of sensor nodes, small size and contingency method. It is also usually not possible to recharge or replace sensor nodes due to the use of these types of networks in harsh and inaccessible environments. Therefore, the most important risk in designing wireless sensor networks is to reduce power consumption in order to increase network lifetime. Therefore, incorporating energy storage algorithms into the design of long-life sensor networks is critical to increase the reliability of the target.

## 2. Introducing the overall structure of the wireless sensor network

In the structure of the wireless sensor network, sensor nodes are used. Despite the different applications for sensor nodes in wireless networks, the primary task of sensor nodes is to sense and collect data and transfer it to the central station. [3] Sensor nodes are the source nodes that transmit data. Due to the inherent limitations of the wireless sensor networks or because of the location of the sensor nodes relative to the central station they cannot directly communicate with the central station via a jump, thus requiring multi-path routes to Are the central station. Our sensor nodes located between the nodes of the central station in addition to performing the assigned tasks should act as intermediaries when transmitting other sensor nodes.

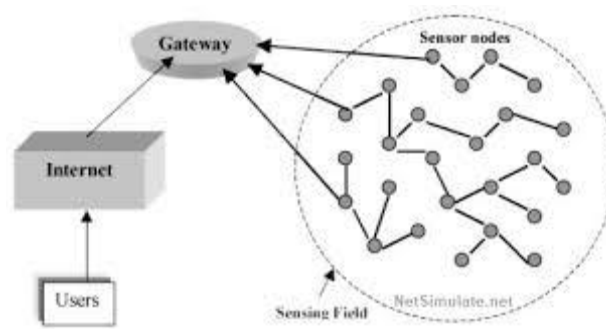

**Figure 1: The overall structure of the wireless sensor network**

## 3. Reliability in wireless sensor networks

Any node may crash or be completely destroyed by environmental events such as a crash or explosion, or it may be disrupted by the power source running out. Tolerance or reliability means that node failure should not affect overall network performance. [4]

## 4. The effect of reduced power consumption on wireless sensor network capability

In wireless sensor networks, when the data is received by a node in one area, that node must forward its received information to all neighboring nodes. If this is repeated several times, the node loses its energy and becomes operational. Failure to interfere with the exchange of network information no longer receives information from that area of the wireless sensor network, which reduces network reliability and lifetime. Because the topology of wireless sensor networks is constantly changing [6] and radio interference also causes a large number of packets to be lost, it is very difficult to guarantee the capability of these networks. Damage reliability and tolerance improve by sending multiple copies of data across multiple paths. Although this method increases power consumption, it will reduce the chance of data loss when links are broken. There are various ways to reduce this overhead caused by data transmission. Another set of protocols use only one path at a time to transmit data [7], and in the event of a major route failure, they will send data over the backup path, thus potentially losing data due to the crash. The path is reduced.

## 5. Introducing a Method to Reduce Power Consumption in Wireless Sensor Networks

MESH structure: Mesh network is widely used due to its simple structure and unique topology properties in many networks, including traditional network and interconnection networks.

As shown in Fig. 2, there is a sensor in the Mesh network to create a wireless sensor network using the Mesh structure instead of each node and the distance of both adjacent nodes (or the edges of the Mesh network) We divide it into sensors in each range of wireless signal transmit power. Of course, all the edges in the Mesh network are equal, but the transmission of information over the network is the same as traditional Mesh networks, and each sensor node is only It can send information to its neighbor nodes in the traditional Mesh network.

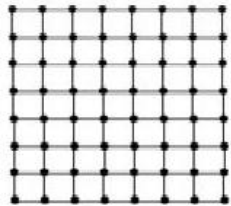

**Figure 2: Mesh structure**

## 6. Describes the LEACH-based clustering method

Since the sensors in the network are under high pressure, it is not possible for all sensors to send information directly to the main station. Information generated from neighboring sensors transmits data seamlessly. In addition, the volume of information generated on broadband networks usually requires more stations. For such problems, we can use data aggregation at sensor nodes. Information aggregation usually involves linking information from multiple sensors to the middle nodes and transmitting the information gathered to the main station. Information aggregation can eliminate extra bits and minimize the number of transfers and ultimately save energy [5 In addition to supporting information gathering through network organization, nodes can be grouped into small groups known as clusters. Each cluster has a header and the other nodes are part of that cluster. Creating clusters and assigning header tasks can play a large role in scalability, lifetime and energy savings of the entire network. LEACH is one of the most well-known algorithms for energy saving classifications in wireless sensor networks that node these headers based on the received signal power and uses these local headers as the main station because they transmit data to The main station consumes a lot of energy [1] Switches all sensor nodes inside a cluster by rotating the header. This results in the power consumption of all the nodes being balanced and hence increasing the lifetime of the network. In contrast to the LEACH algorithm, another algorithm called

LEACH-C is proposed. This algorithm is a kind of centralized clustering algorithm that is similar to the LEACH steady state. In LEACH-C, each node sends information about the current location and amount of energy left to the main station. In addition to determining good clusters, the main station must ensure even distribution of energy loads between the nodes. To do this, the main station has to calculate the average energy of the node, and any node that has less energy than this average cannot be the header of the field. LEACH-C does not support the scalability of wireless sensor networks because this requires that the base station have to solve the problem of optimal clustering that is not possible for a network of thousands of nodes. The approach proposed to reduce the residual energy of the node is to modify the beam input which increases the lifetime of the LEACH network by 30% [2] The change in input by residual energy may cause other problems because after several times Repeat this procedure with the remaining nodes having a low energy level and the initial value of the header will be reduced and many headers will not have enough power to transmit information to the main station even though sufficient energy nodes are still available for this purpose. But the network can't work well. The input equivalent can be changed by having an input that increases the input of each node, but is not part of the header for a given number of cycles. This node is more likely to become a header due to higher inputs. In this algorithm, the cluster organization starts in the first step and is followed by a persistence state when transferring data from the node to the main station. During the organizing phase each node selects a random number between 0 and 1, if this number is less than the input, the node becomes the header for the current cycle. The input is organized in such a way that the other nodes choose their own more appropriate header. Are organized into adjacent clusters. During shelf life, the headers collect information from the member nodes and then send the collected information to the main station. Figure 3

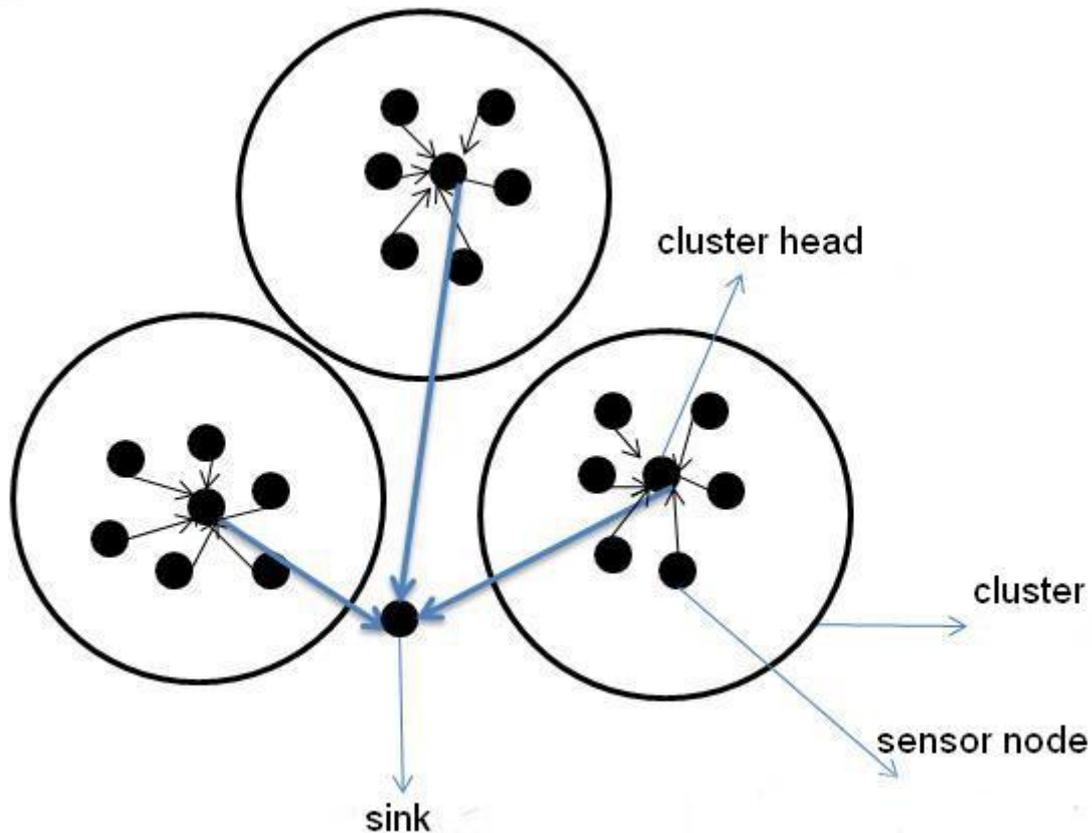

**Figure 3: Sending information in LEACH**

## 7. Modify the proposed algorithm

In the LEACH algorithm, some nodes have to choose a header that is longer than the main station. These sensor nodes send their information to a farther place and then have to travel longer to reach the main station. Such transfers, called redundant transfers, waste network energy. As shown in Figure 4, nodes A and B use the additional transmissions to forward their information to the main station, while nodes C and D use their information. They send to the main station via the main routes without sending to remote locations. In addition to solving the additional transfer problem, we are making a change in the organization stage of the LEACH algorithm.

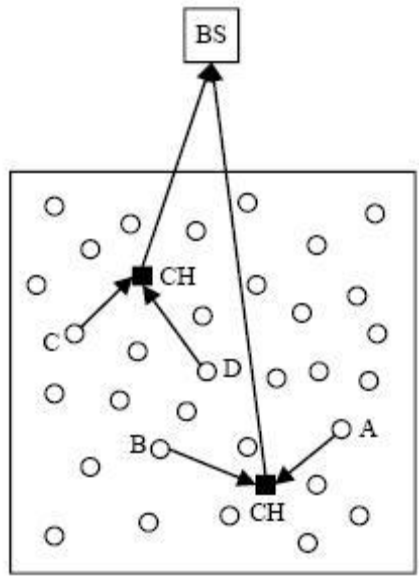

**Figure 4: Additional transitions in the initial LEACH algorithm for nodes A and B**

At this point again, the selected header and the other sensor nodes do not need to select the closest node. Among the headers that are less distant from the main station than their own, these nodes select the nearest header and select it for the member. Clustering declares that if no such header is present, each cluster will become a member and send its related information directly to the main station, as shown in Figure 5. CH1 is the closest header to the node A is node but because node A is longer distance from the main station than node A is not node A Also, since there is no header in the network that is close to the main station compared to B, this node will not be a member of any of the clusters will send its information directly to the main station. [2]

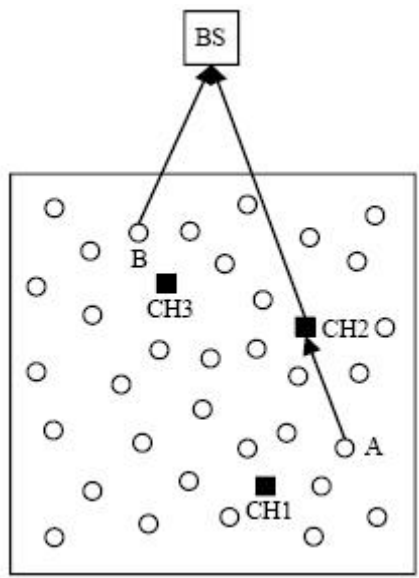

**Figure 5: Transfer of information to the main station in the proposed algorithm**

## 8. Comparison of Proposed Algorithm with Prior Method

Disadvantages of MESH: The high cost of this network due to the large number of nodes and the use of the traditional method and if the node that wants to send the information has trouble getting to the central station the node will never know the information. Sending all the nformation to all neighboring nodes reduces network lifetime. And also compared to the original LEACH algorithm that randomly selects the header, which in the proposed algorithm olves this problem which increases the network lifetime. Has been. Comparison of the proposed algorithm in terms of network lifetime with the previous method in Figure 6 (simulation with Excel software according to the number of sensors)

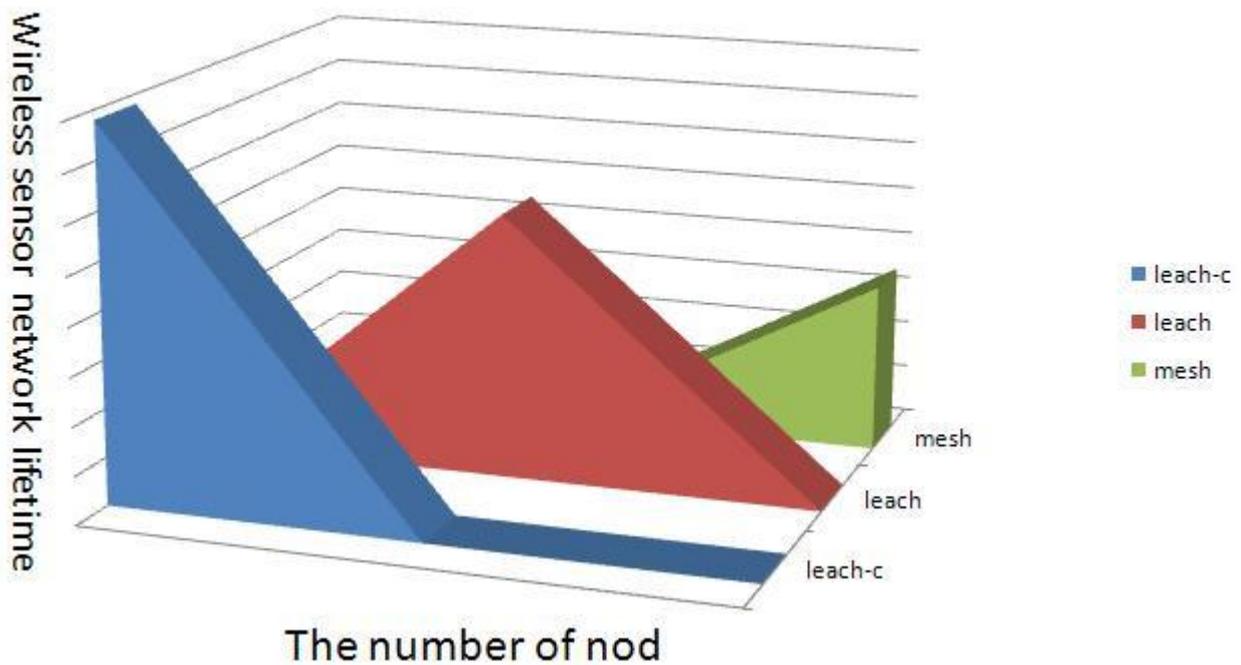

**Figure 6: Comparison in terms of lifetime**

Comparison of the proposed algorithm in terms of network reliability with the previous method in Figure 7

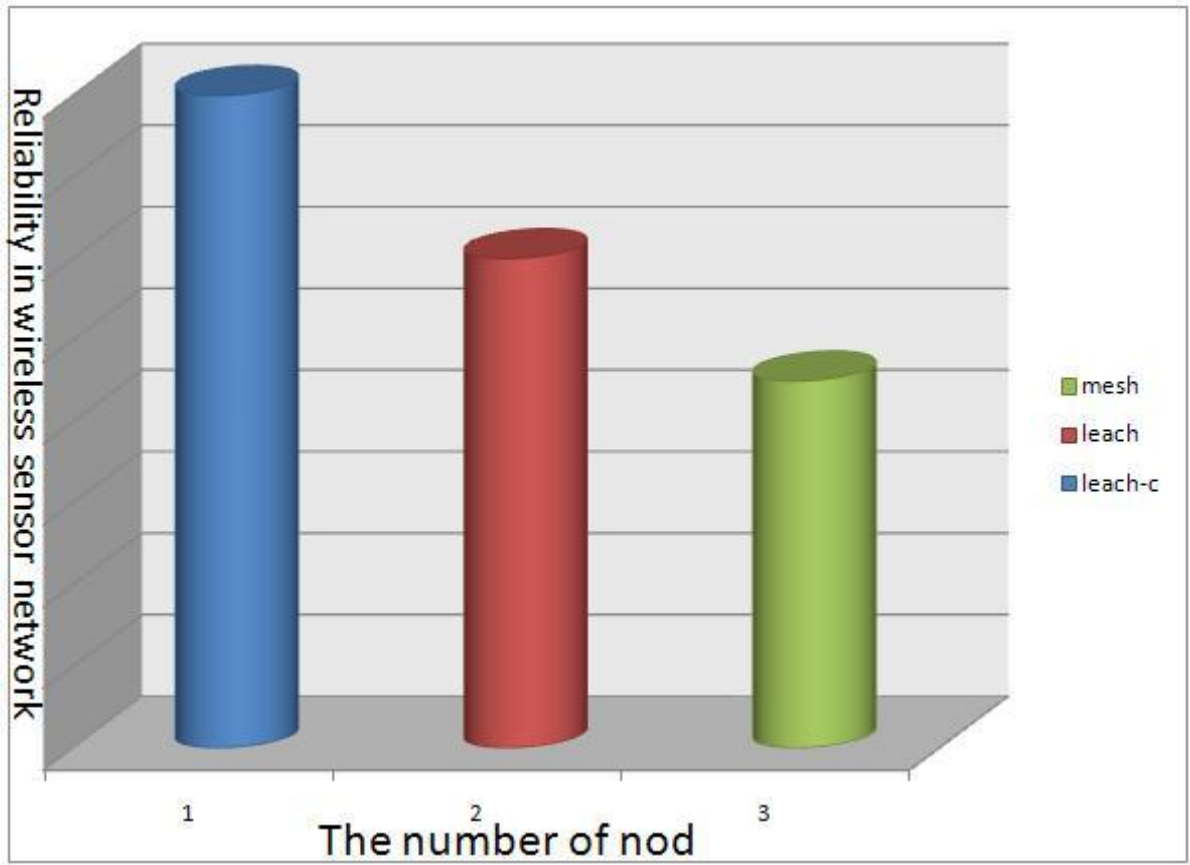

**Figure 7: Reliability comparison**

## 9. Conclusion

In this paper, an efficient LEACH-based clustering method is introduced that reduces the energy consumption of the wireless sensor network and increases the lifetime and reliability of the network, taking into account simulations of this algorithm and over 16% of the total energy. Saves consumption resulting from the elimination of redundant transfers. In the future we can design algorithms to improve the status of nodes around the main station.

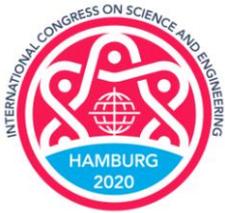
3rd .International Congress on Science and Engineering
HAMBURG – GERMANY
March 2020